\def\year{2015}
\documentclass[letterpaper]{article}
\usepackage{aaai}
\usepackage{graphicx}
\usepackage{amsmath}
\usepackage{amssymb}
\usepackage{amsfonts}
\usepackage{float}
\usepackage{url}
\usepackage{xspace}
\usepackage{hyperref}
\usepackage{ifpdf}
\usepackage{xcolor}
\usepackage{cite}
\usepackage{natbib}
\newif\ifshowcomments
\showcommentstrue

\ifshowcomments
\newcommand{\mynote}[2]{\fbox{\bfseries\sffamily\scriptsize{#1}}
 {\small$\blacktriangleright$\textsf{\emph{#2}}$\blacktriangleleft$}}
\else
\newcommand{\mynote}[2]{}
\fi

\begin{document}
%
\title{Everyday the Same Picture: Popularity and Content Diversity}

\author{Alessandro Bessi \\IUSS Pavia, Italy \And
Fabiana Zollo \\ IMT Lucca, Italy \And
Michela Del Vicario \\ IMT Lucca, Italy \AND  
Antonio Scala \\ ISC CNR,Rome, Italy \And
Fabio Petroni \\ Sapienza University of Rome, Italy \AND
Bruno Gon\c calves \\ Aix Marseille, France \And
Walter Quattrociocchi \\ IMT Lucca, Italy \\ \textbf{walter.quattrociocchi@imtlucca.it}
}
\maketitle

\begin{abstract}
Facebook is flooded by diverse and heterogeneous content, from kittens up to music and news, passing through satirical and funny stories. 
Each piece of that corpus reflects the heterogeneity of the underlying social background.
In the Italian Facebook we have found an interesting case: a page having more than $40K$ followers that every day posts the same picture of a popular Italian singer.
In this work, we use such a page as a control to study and model the relationship between content heterogeneity on popularity. 
In particular, we use that page for a comparative analysis of information consumption patterns with respect to pages posting science and conspiracy news. 
In total, we analyze about $2M$ likes and $190K$ comments, made by approximately $340K$ and $65K$ users, respectively. 
We conclude the paper by introducing a model mimicking users selection preferences accounting for the heterogeneity of contents.
\end{abstract}


\section{Introduction}
Online social networks such as Facebook foster the aggregation of people around common interests, narratives, and worldviews. Indeed, the World Wide Web caused a shift of paradigm in the production and consumption of contents that increased volume and heterogeneity of available contents. Users can express their attitudes by producing and consuming heterogeneous information --- e.g. conspiracists avoid mainstream news and follow their own information sources, whereas debunkers try to inhibit the diffusion of false claims. 
Images of kittens and pets, political memes, gossip, scandals spread on Facebook. 
By liking, commenting, and sharing their preferred contents, users can express their passions and emotions --- and, among these latter, sarcasm in not an exception. 
Indeed, not rarely we can find pages promoting parodistic and sarcastic imitations of online social dynamics --- e.g., {\em Ebola and Kittens} \cite{kittens} or {\em In favor of chem-trails} \cite{chemtrails}.
An interesting case in the Italian Facebook is a page \cite{TotoCutugno} with more than $40K$ followers that posts everyday the exactly alike picture  of Toto Cutugno, a famous Italian pop-singer.

In this work, we use the intriguing case of that page as a baseline to study and model the effect of content diversity on popularity. 
Specifically, we analyze user activity and post consumption patterns on the baseline page for a timespan of about $4$ months. 
Through a comparative analysis between two sets of pages producing heterogeneous contents, we show that there are no remarkable differences in user activity patterns, whereas significant dissimilarities between post consumption patterns emerge. 
Such a comparative analysis allows to derive a model of information consumption accounting for the heterogeneity of contents. 
Hence, we show that the proposed model is able to reproduce the phenomenon observed from empirical data. In particular, we show the effects of different levels of contents' heterogeneity on posts consumption patterns.  


\section{Background and Related Works}
A large body of literature addresses the study of social dynamics on socio-technical systems from social contagion up to social reinforcement \cite{Onnela2010,Ugander2012,Lewis2012,Mocanu2012,Adamic2005,Kleinberg2013,eRep,Quattrociocchi2011,Quattrociocchi2009,Bond2012,Moreno2011,Centola2010,Castellano2007,Quattrociocchi2014,Bennaim2003,Adamic2013,Hannak2012,Cheng2014, Goncalves2012}.
Among these, one of the most defining topic of computational social science is the understanding of driving forces behind the popularity of contents \cite{tatar2014survey}.
Such a challenge has been addressed looking at the sentiment of comments, contents, or users' attention \cite{tatar2011predicting,figueiredo2014does,zadeh2014modeling,ratkiewicz2011detecting,bandari2012pulse,gomez2008statistical,szabo2010predicting,lerman2010using,lerman2010information}.
However, the mechanisms behind popularity remain largely unexplored \cite{Goldhaber1997,Watts2002,Leskovec2009}. 
In \cite{salganik2006experimental} the authors address such a challenge experimentally by measuring the
impact of content quality and social influence on the eventual popularity or success of cultural artifacts.
The effects of specific contents on the formation of communities of interest, their permeability to false information, and the resistance to changes have been recently characterized in \cite{JTM2014,Rojecki2014,SOCINFO2014, PLOS2014}. In particular, in \cite{WWW2015} the authors point out that connectivity patterns of the Facebook social network are prominently driven by homophily of users --- i.e., the tendency of individuals to associate with similar others --- towards specific kinds of contents. Microblogging platforms such as Facebook and Twitter \cite{Tapscott2006} have lowered the cost of information production and broadcasting, boosting the potential reach of each idea or meme \cite{Dawkins1989,Bauckhage2011} --- i.e., content or concepts that spread rapidly on the Web. Still, the abundance of information to which we are exposed through online social networks and other socio-technical systems is exceeding our capacity to consume it \cite{Weng2012}. 
As a result, the dynamics of information is driven more than ever before by the economy of attention \cite{Simon1971,dukas2001limited, Lehmann2012}.
We address this challenge by studying the interlink between contents diversity and popularity. More specifically, we investigate the effects of sources producing always the same information on users' activity, consumption, and attention patterns.

\section{Data Description}
In this work, we aim at investigating the role of content diversity on the dynamics of information consumption in online social networks. 
To this end, we use a set of Facebook pages promoting heterogeneous contents and a Facebook page promoting always the same picture. The set of pages promoting heterogeneous contents is composed by $73$ public Facebook pages, whereof $34$ are about scientific news and $39$ about conspiratorial news; we refer to the former as \emph{science pages} and to the latter as \emph{conspiracy pages}. The page promoting homogeneous contents is called "La stessa foto di Toto Cutugno ogni giorno" ("Everyday the same photo of Toto Cutugno", a well-known Italian pop singer-songwriter); such a page, by publishing everyday the same picture of the Italian singer --- and nothing else ---, represents the perfect control for studying content diversity; we refer to this page as the \emph{baseline page}.
Starting from these pages, we downloaded all the posts, we collected all the \emph{likes} and \emph{comments} to the posts, and we counted the number of \emph{shares}. Data related to science and conspiracy pages have been collected from August 22, 2013 to December 31, 2013, whereas data related to the baseline page have been collected from August 22, 2014 (birthdate of the page) to December 31, 2014. 
In total, we collected around $2M$ likes and $190K$ comments, made by about $340K$ and $65K$ users, respectively. 
In Table \ref{tab:1} we summarize the details of our data collection.
Likes, shares, and comments have a different meaning from the user viewpoint. Most of the times, a like stands for a positive feedback to the post; a share expresses the will to increase the visibility of a given information; and a comment is the way in which online collective debates take form. Comments may contain negative or positive feedbacks with respect to the post.

\begin{center}
\begin{table}[!h]
\centering
\tiny
\begin{tabular}{l|c|c|c|c}
\hline\bf { }  & \bf {Total} & \bf {Science} & \bf {Conspiracy} & \bf {Baseline} \\ \hline 
Pages & $ 74 $ & $ 34 $ & $ 39 $ & $1$ \\
Posts & $ 49,354 $ & $ 13,028 $ & $ 36,169 $ & $157$ \\
Likes & $ 2,095,677 $ & $ 614,078 $ & $ 1,184,084$ & $297,515$   \\
Comments & $192,967 $ & $ 40,608 $ & $ 138,138 $ & $14,221$ \\
Shares &  $3,782,480$ & $477,457$ & $3,297,687$ & $7,336$ \\
Likers & $ 344,367 $ & $ 162,146 $ & $ 159,524 $ & $22,697$ \\
Commenters & $ 64,903 $ & $ 18,358 $ & $ 41,666 $ & $4,875$ \\
\end{tabular}\newline
  \caption{ \textbf{Dataset breakdown.} The number of pages,
    posts, likes, comments, shares, likers, and commenters for science pages, conspiracy pages, and the baseline page.}
  \label{tab:1}
\end{table}
\end{center}

\section{Results and Discussion}
In this section, we first present the statistical signatures characterizing users activity on pages with diversified content on specific topics (science and conspiracy news) against the case of the page posting every day the same picture (baseline). Then we derive a model of information consumption mimicking user preferences with respect to contents.

\subsection{Content and Users Activity}
\begin{figure}[h]
	\centering\includegraphics[width = 0.45\textwidth]{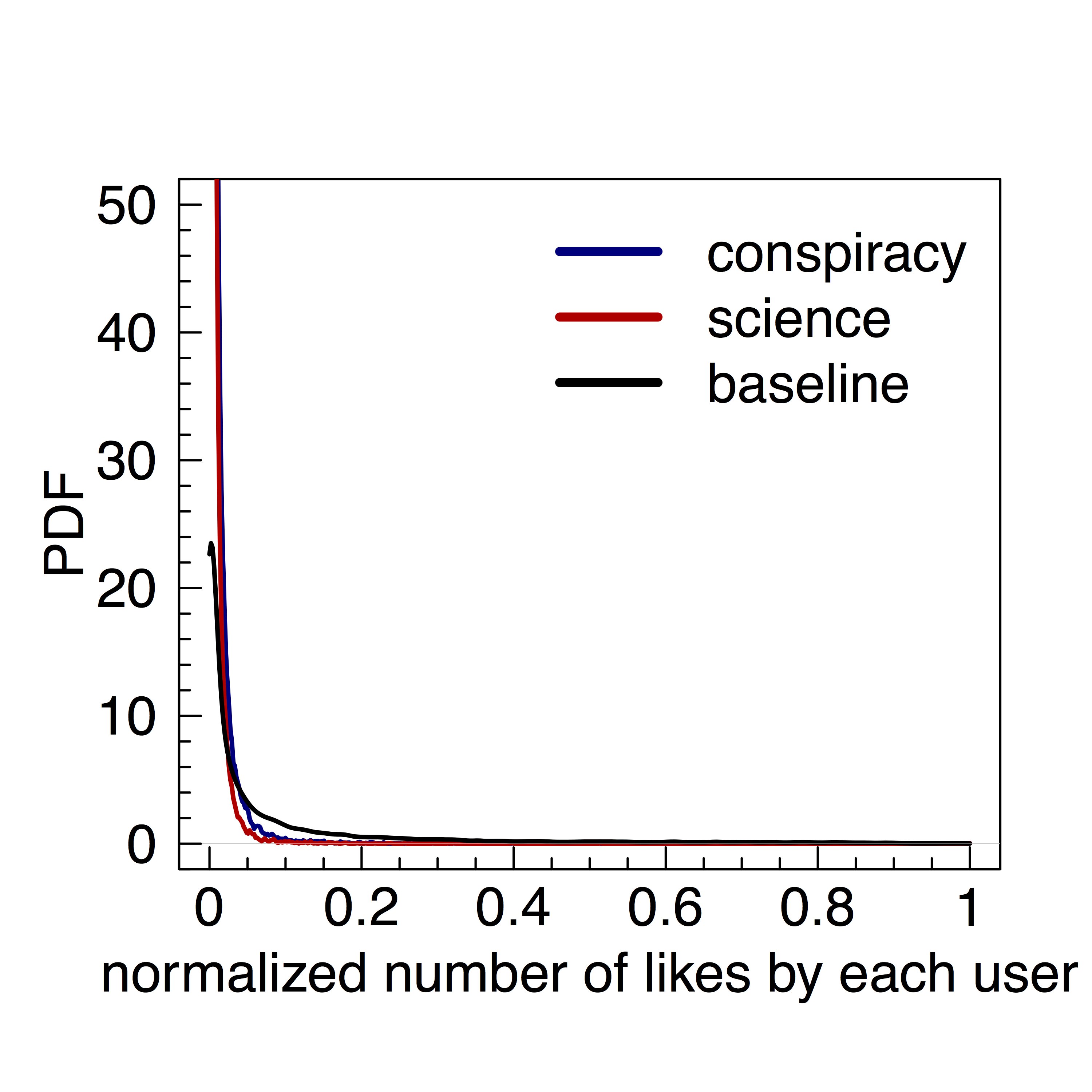}
	\caption{\textbf{Users' activity patterns.} Probability density function (PDF) for the normalized number of likes by each user.}
	\label{fig:2}
\end{figure}

Let us focus some regularities concerning users' activity on science pages and conspiracy pages compared with the baseline page. 
Figure \ref{fig:2} shows the probability density function (PDF) for the normalized\footnote{We performed the unity--based normalization to bring all values in the range $[0,1]$.} number of likes for each user. We find that the activity of users presents an heavy--tailed distribution. 

\begin{figure}[h]
	\centering\includegraphics[width = 0.45\textwidth]{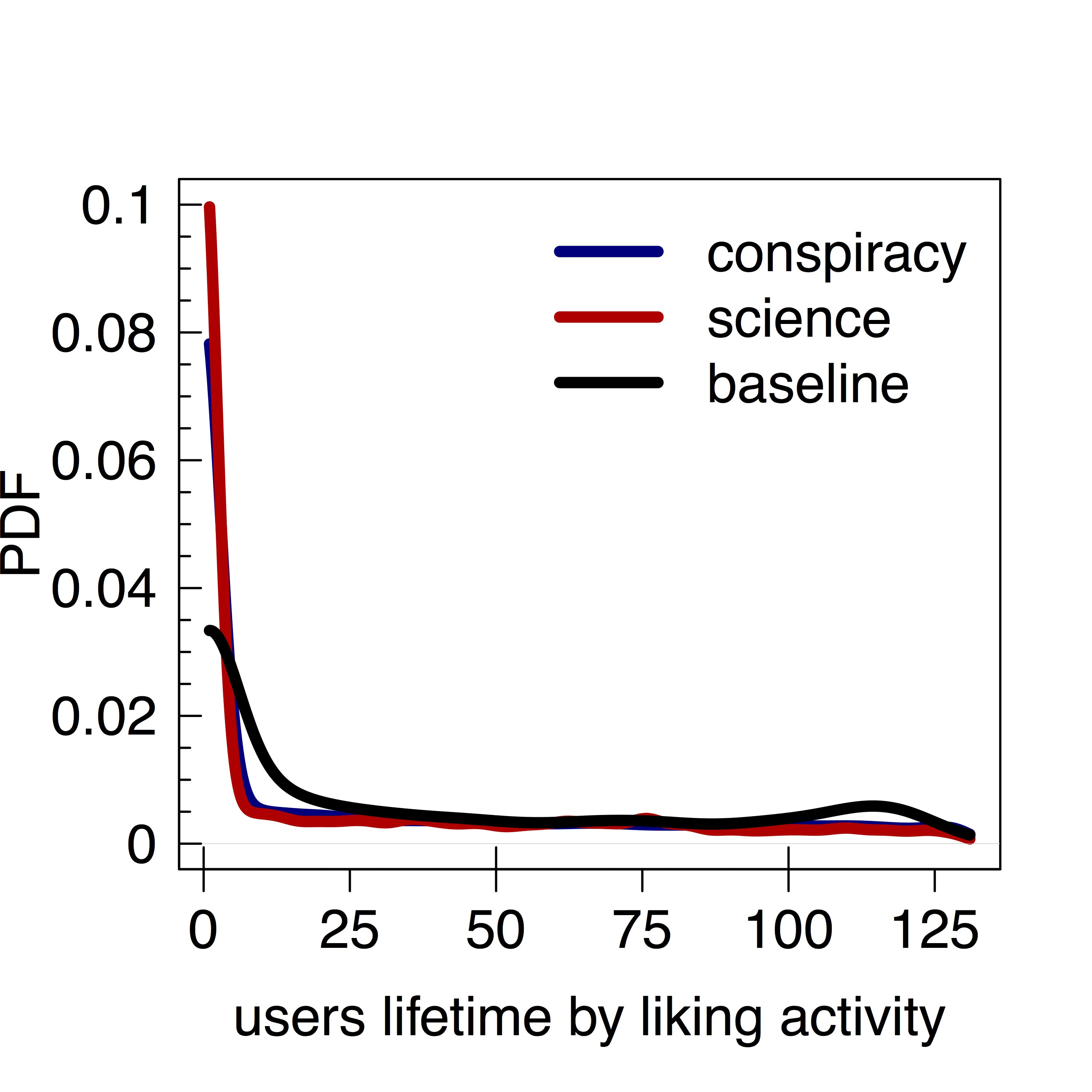}
	\caption{\textbf{Users' lifetime.} Probability density function (PDF) of the users' lifetime in terms of their liking activity. The PDF shows a slight difference in the lifetime of the baseline users with respect to science and conspiracy users. 
		}
	\label{fig:3}
\end{figure}

In Figure \ref{fig:3} we show the PDF of the users' lifetime in terms of their liking activity --- i.e. the temporal interval between the first and the last like of the user on a given page. We find a slight difference in the lifetime of the baseline users with respect to science and conspiracy users. 

 These figures show that users activity patterns are similar and present heavy--tailed distributions despite the different nature of the contents, and we can not find any significant difference between the users interaction patterns induced by heterogeneous or homogeneous contents.
 
\begin{figure}[h]
	\centering\includegraphics[width = 0.45\textwidth]{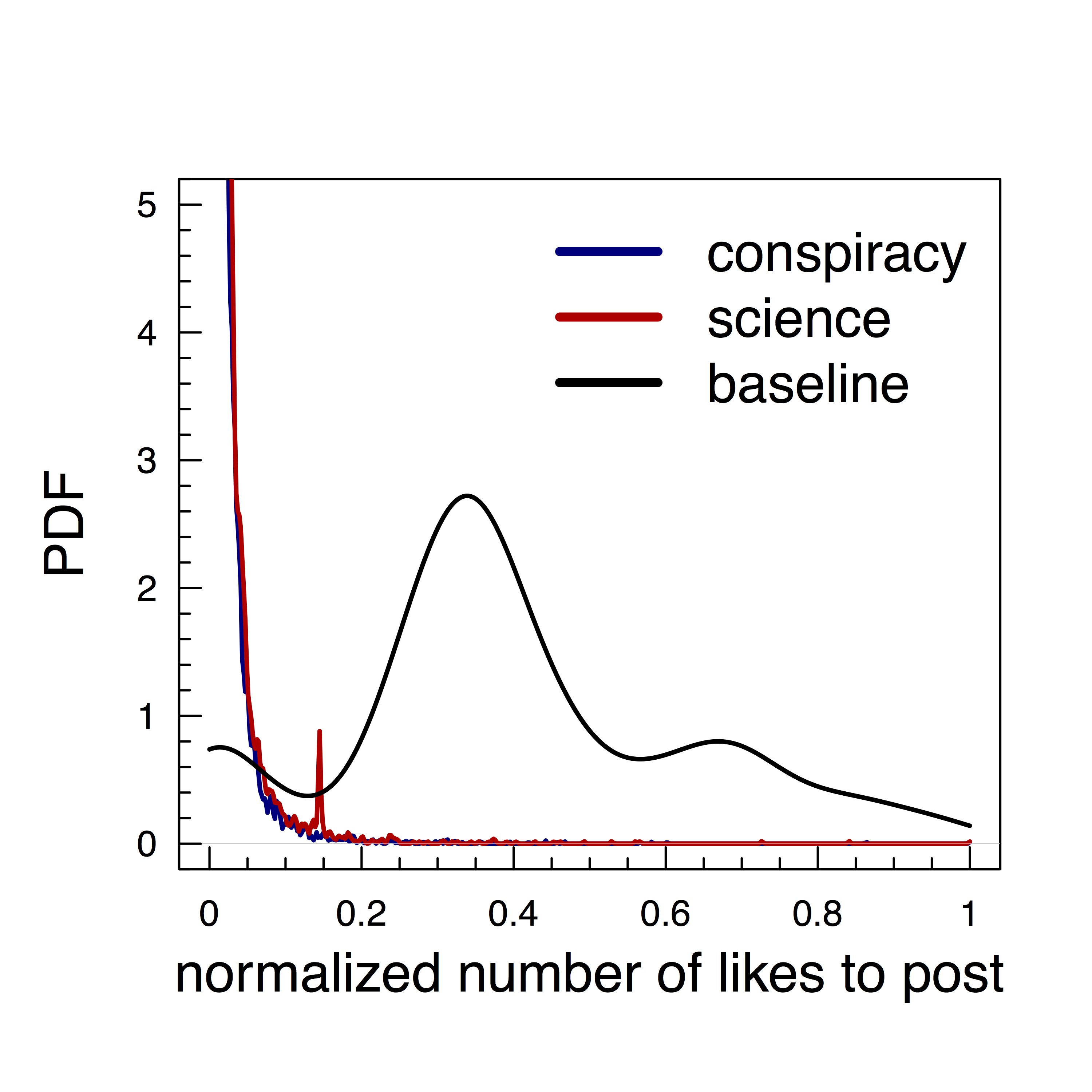}
	\caption{\textbf{Posts' consumption patterns.} Probability distribution function (PDF) for the normalized number of likes received by posts belonging to science pages, conspiracy pages, and the baseline page. The PDFs show remarkable differences between consumption patterns' distributions related to pages promoting heterogeneous contents and those related to the page promoting homogeneous contents.}
	\label{fig:5}
\end{figure}

Conversely, by analyzing consumption patterns related to posts, we find a significant difference in the information consumption dynamics. Figure \ref{fig:5} shows the PDF for the number of likes received by posts belonging to science pages, conspiracy pages, and the baseline page.
The number of likes received by posts are heavy--tailed distributed if the posts belong to pages promoting heterogeneous contents (science and conspiracy pages); whereas they are approximately distributed according to a Gaussian if the posts belong to a page promoting homogeneous content (baseline page). 

Summarizing, users' activities always present heavy--tailed distributions resolving in heavy--tailed distributed consumption patterns on posts in the heterogeneous contents case.  Still, when the content promoted by a page is homogeneous -- i.e., always the same -- we find that the heavy--tailed distributed users' activities resolve in posts' consumption patterns that are approximately Gaussian.

\subsection{Modeling Contents Consumption}
Here we introduce a model of pattern consumption that exploits the Beta distribution properties to generate different levels of posts' attractiveness, thus varying content--heterogeneity in the simulated collection of posts.
 
\begin{figure}[h]
	\centering\includegraphics[width = 0.45\textwidth]{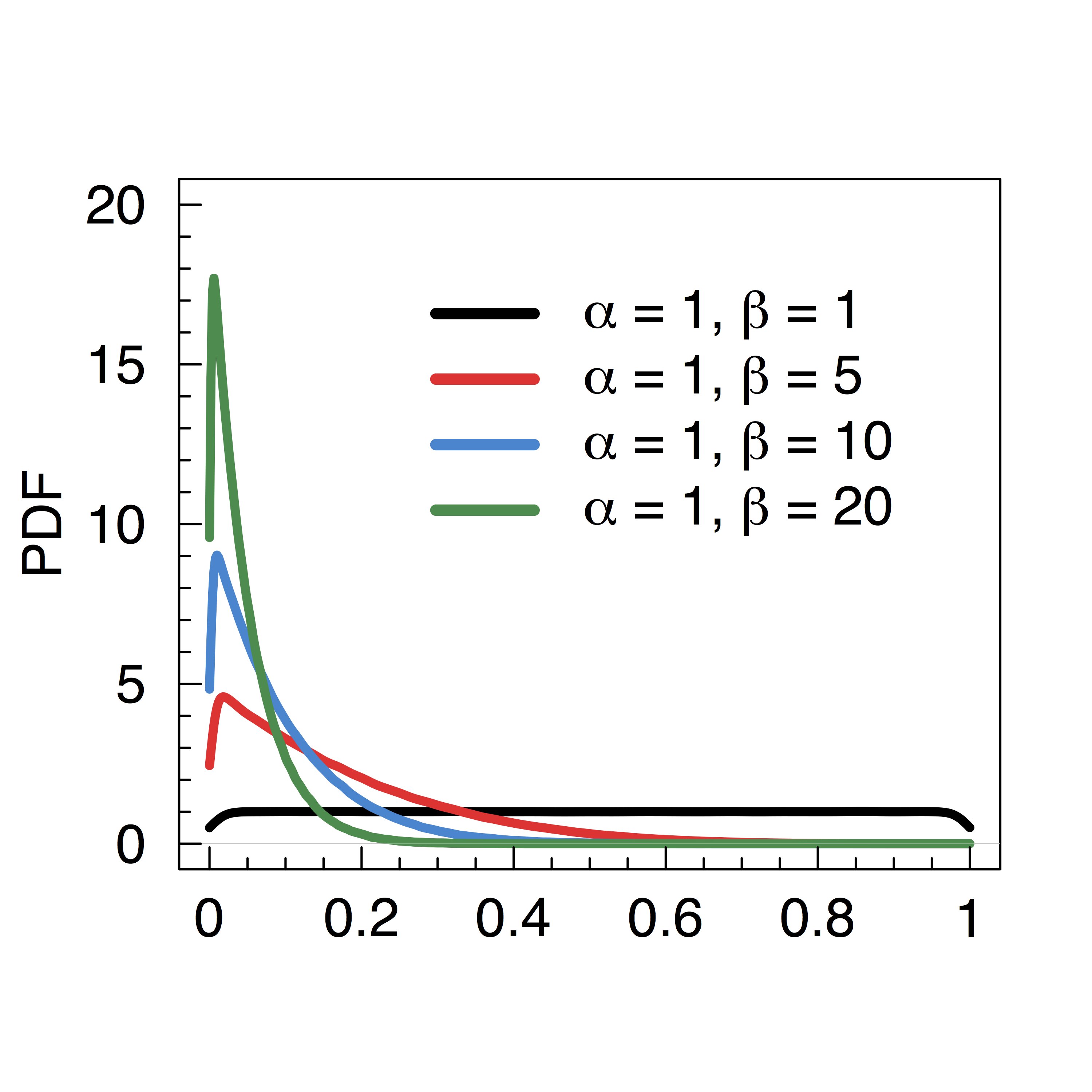}
	\caption{\textbf{Beta distribution $\mathcal{B}e(\alpha,\beta)$.} Two parameters, $\alpha$ and $\beta$, control the shape of the distribution. In particular, for $\alpha = 1$ and $\beta = 1$ the Beta distribution $\mathcal{B}e(\alpha,\beta)$ is equivalent to the Uniform distribution $\mathcal{U}(0,1)$. Conversely, if $\alpha = 1$ and $\beta \gtrsim 20$, the Beta distribution $\mathcal{B}e(\alpha,\beta)$ is a right heavy--tailed distribution.}
	\label{fig:1}
\end{figure}

The Beta distribution is a family of continuous probability distributions defined in the interval $[0,1]$ and characterized by two real parameters, $\alpha > 0$ and $\beta > 0$, which control the shape of the distribution. In particular, for $\alpha = 1$ and $\beta = 1$ the Beta distribution $\mathcal{B}e(\alpha,\beta)$ is equivalent to the Uniform distribution $\mathcal{U}(0,1)$. 
Conversely, if $\alpha = 1$ and $\beta \gtrsim 20$, the Beta distribution $\mathcal{B}e(\alpha,\beta)$ is a right heavy--tailed distribution. Figure \ref{fig:1} shows the Beta probability density function with respect to the two shape parameters $\alpha$ and $\beta$.

In our model, each post has a value drawn from a Beta distribution $v \sim \mathcal{B}e(1, \beta)$, with $\beta$ ranging between $1$ and $1,000,000$, indicating its attractiveness. We let the parameter $\beta$ assume those extreme values in order to obtain different distributions for posts' attractiveness.
Indeed, notice that when $\beta = 1$ the Beta distribution $\mathcal{B}e(1, \beta)$  is equivalent to a uniform distribution $\mathcal{U}(0,1)$, so that we have a collection of homogeneous--content posts --- i.e., each post has the same degree of attractiveness; whereas when $\beta \to \infty$ the Beta distribution $\mathcal{B}e(1, \beta)$ is equivalent to a right heavy--tailed distribution, so that we have a collection of heterogeneous--content posts --- i.e., there are few posts with a high level of attractiveness, while the vast majority of the posts is characterized by a low level of attractiveness. Moreover, each user is characterized by two parameters randomly drawn from power law distributions: her volume of activity, $a \sim p(x)$; and her fixed--preference about the posts, $b \sim p(x)$, where $p(x) = x^{-\gamma}$ with $\gamma = 1.5$. 
Each user can not exceed her assigned volume of activity, $a$, and she likes a given post if and only if her normalized\footnote{Note that we performed a unity--based normalization in order to bring all values of $b \sim p(x) = x^{-1.5}$ in the range $[0,1]$, so that the fixed--preference of the user is comparable with the attractiveness of the posts.} fixed--preference, $b$, is smaller than the attractiveness, $v$, of that post. Note that in our model we do not take into account the users' network: since Facebook network is very dense --- indeed, the diameter of Facebook social network is $3.74$ \cite{Backstrom2012,WWW2015} --- the connections between users are not likely to influence posts' consumption dynamics.

We run simulations for $\beta$ ranging between $1$ and $1,000,000$, with $P = 10,000$ (posts) and $U = 20,000$ (users). Results are averaged over $100$ iterations.

%

\begin{figure*}
	\centering\includegraphics[width = 0.85\textwidth]{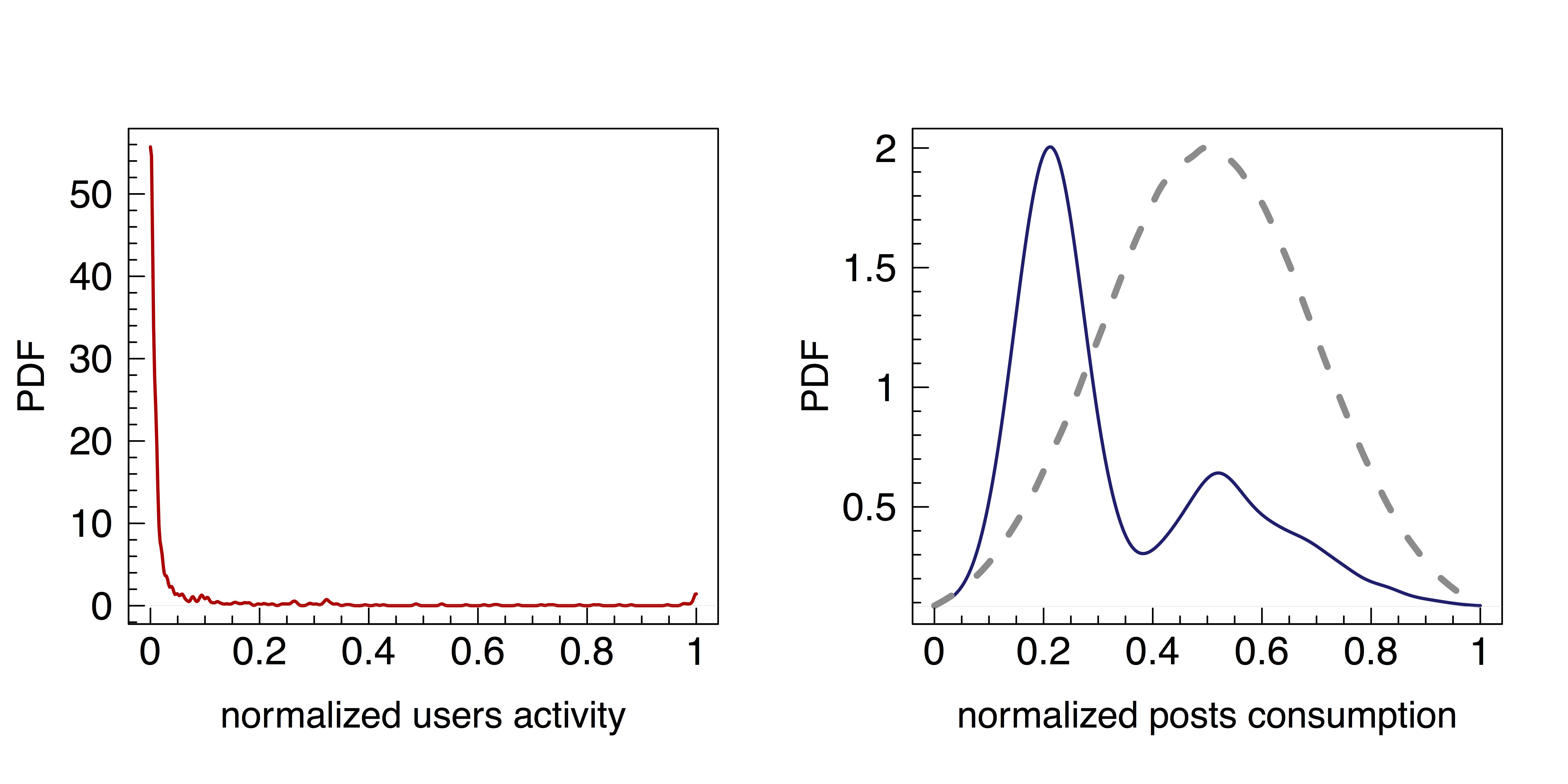}
	\caption{\textbf{Users activity and post consumption patterns with extremely heterogenous--content posts.} Probability density function (PDF) of the users activity and the posts consumption patterns generated by a simulation of the model with $\beta = 1,000,000$. If the content promoted by a page is heterogeneous, the heavy--tailed users' activity resolves in skewed posts consumption's patterns.}
	\label{fig:7}
\end{figure*}

Figure \ref{fig:7} shows the probability density function (PDF) of the users activity and the posts consumption patterns generated by a simulation of the model with $\beta = 1,000,000$ --- i.e., in the case of extremely heterogeneous--content posts. Observe that users' activity is heavy--tailed, and the distribution of posts' consumption is skewed. Such a result is consistent with empirical data shown in the previous section: if the content promoted by a page is heterogeneous, the heavy--tailed users' activity resolves in skewed posts consumption's patterns.

\begin{figure*}
	\centering\includegraphics[width = 0.85\textwidth]{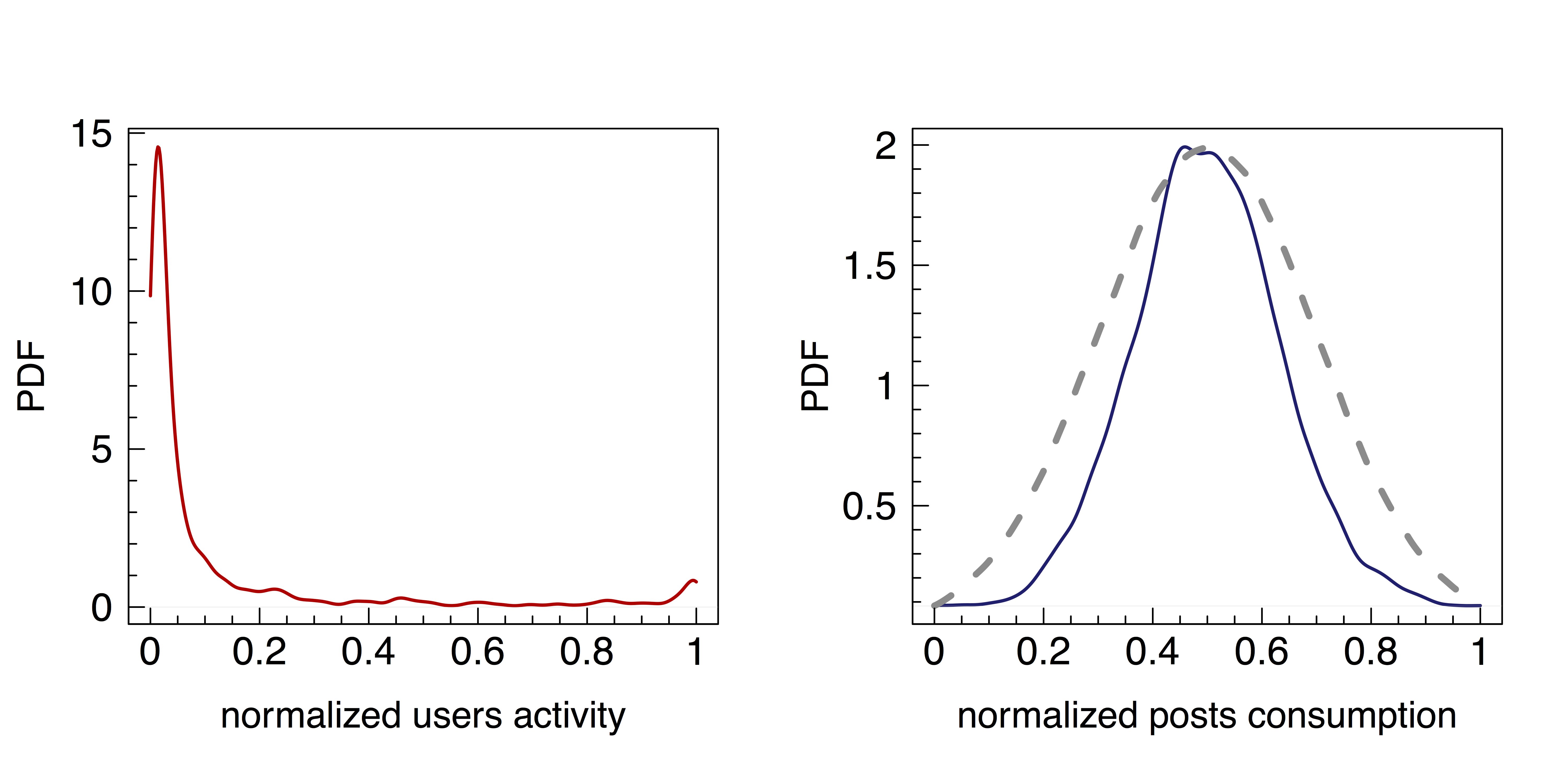}
	\caption{\textbf{Users activity and post consumption patterns with homogeneous--content posts.} Probability density function (PDF) of the users activity and the posts consumption patterns generated by a simulation of the model with $\beta = 1$. If the content promoted by a page is always the same, the heavy--tailed users' activity resolves in approximately Gaussian posts consumption's patterns.}
	\label{fig:8}
\end{figure*}

Figure \ref{fig:8} shows the probability density function (PDF) of the users activity and the posts consumption patterns generated by a simulation of the model with $\beta = 1$ --- i.e., in the case of homogeneous--content posts. Notice that users' activity is heavy--tailed, whereas posts' consumption is approximately Gaussian. Such a result is consistent with empirical data shown in the previous section: if the content promoted by a page is always the same, the heavy--tailed users' activity resolves in approximately Gaussian posts consumption's patterns.

\section{Concluding Remarks}
Facebook is overflowed by different and heterogeneous contents, from the latest news up to satirical and funny stories. Each piece of that corpus reflects the heterogeneity of the underlying social background. Indeed, the World Wide Web caused a shift of paradigm in the production and consumption of information that increased the amount and heterogeneity of contents available to users. On online social networks such as Twitter and Facebook, people can express their attitudes, passions, and emotions by producing and consuming heterogeneous information.

In the Italian Facebook, we have found a fascinating case of contents' homogeneity: a page with more than $40K$ followers that every day posts the same picture of Toto Cutugno, a popular Italian singer. In this work, we use such a page as a benchmark to investigate and model the effect of contents’ heterogeneity on popularity. In particular, we use that page for a comparative analysis of information consumption patterns with respect to pages posting heterogeneous contents related to science and conspiracy.

We show that there are not remarkable differences in user activity patterns, whereas we find significant dissimilarities between post consumption patterns of the page promoting homogeneous contents and those of the pages producing heterogeneous contents. Finally, we derive a model of information consumption that accounts for the heterogeneity of contents. Hence, we show that the proposed model is able to reproduce the phenomenon observed
from empirical data.

\section{Acknowledgments}
Funding for this work was provided by EU FET project MULTIPLEX nr. 317532 and SIMPOL nr. 610704. The funders had no role in study design, data collection and analysis, decision to publish, or preparation of the manuscript.
We want to thank Prof. Guido Caldarelli for precious insights and contribution on the data analysis.
Special thanks to Josif Stalin, Stefano Alpi, Michele Degani for giving access to the Facebook page of {\em La stessa foto di Toto Cutugno ogni giorno}.
	
\bibliographystyle{unsrt}
\bibliography{toto}

\end{document}